\documentclass[14pt, preprint2]{aastex6}
\usepackage{bm, graphicx, subfigure, amsmath, morefloats, mathtools, amsfonts}
\hypersetup{allcolors = [rgb]{0., 0., 0.5}} 

\usepackage{longtable}
\newcommand{\project}[1]{\textsl{#1}}
 
\newcommand{\tc}{\project{The~Cannon}} 
\newcommand{\apogee}{\project{\textsc{apogee}}}

\newcommand{\documentname}

\newcommand{\teff}{\mbox{$T_{\rm eff}$}}
\newcommand{\tlag}{\mbox{$T_{\rm lag}$}}

\newcommand{\numax}{\mbox{$\rm \nu_{max}$}}
\newcommand{\deltanu}{\mbox{$\rm \Delta{\nu}$}}

\newcommand{\feh}{\mbox{$\rm [Fe/H]$}}

\newcommand{\logg}{\mbox{$\log g$}}

\newcommand{\set}[1]{\bm{#1}}
\newcommand{\starlabel}{\ell}
\newcommand{\starlabelvec}{\set{\starlabel}}

\newcommand{\given}{\,|\,}

\newcommand{\xone}{\ensuremath{x_n}}
\newcommand{\xtwo}{\ensuremath{x_{n'}}}


\begin{document}
\title{Inference of stellar parameters from brightness variations}

\author{Melissa K.~Ness\altaffilmark{1,2}, Victor Silva Aguirre\altaffilmark{3}, Mikkel N.~Lund\altaffilmark{3}, Matteo Cantiello\altaffilmark{2}, Daniel Foreman-Mackey\altaffilmark{2}, David~~W.~~Hogg\altaffilmark{2,4,5}, Ruth Angus\altaffilmark{1}}

\altaffiltext{1}{Department of Astronomy, Columbia University, Pupin Physics Laboratories, New York, NY 10027, USA}\
\altaffiltext{2}{Center for Computational Astrophysics, Flatiron Institute, 162 Fifth Avenue, New York, NY 10010, USA}
\altaffiltext{3}{Stellar Astrophysics Centre, Department of Physics and Astronomy, Aarhus University, Ny Munkegade 120, DK-8000 Aarhus C, Denmark}
\altaffiltext{4}{Center for Data Science, New York University, 60 5th Avenue, New York, NY 10011, USA}
\altaffiltext{5}{Center for Cosmology and Particle Physics, Department of Physics, New York University, 726 Broadway, New York, NY 10003}

\email{melissa.ness@columbia.edu}

\begin{abstract} 

It has been demonstrated that the time variability of a star's brightness at different frequencies can be used to infer its surface gravity, radius, mass, and age. With large samples of light curves now available from \textsl{Kepler} and K2, and upcoming surveys like \textsl{TESS}, we wish to quantify the overall information content of this data and identify where the information resides. As a first look into this question we ask which stellar parameters we can predict from the brightness variations in red-giant stars data and to what precision, using a data-driven, non-parametric model.
We demonstrate that the long-cadence (30-minute) Kepler light curves for 2000 red-giant stars can be used to predict their \teff\ and \logg.
Our inference makes use of a data-driven model of a part of the autocorrelation function (ACF) of the light curve, where we posit a polynomial relationship between stellar parameters and the ACF pixel values.
We find that this model, trained using 1000 stars, can be used to recover the temperature \teff\ to $<100$~K, the surface gravity to $<0.1$~dex, and the asteroseismic power-spectrum parameters \numax\ and \deltanu\ to $<11$~$\mu$Hz and $<0.9$~$\mu$Hz {($\lesssim$ 15\%)}.
We recover \teff\ from range of time-lags 0.045 $<$ \tlag\  $<$ 370 days and the \logg, \numax\ and \deltanu\ from the range 0.045 $<$ \tlag\ $<$ 35 days.
We do not discover any information about stellar metallicity in this model of the ACF.
The information content of the data about each parameter is empirically quantified using this method, enabling comparisons to theoretical expectations about convective granulation. 
\end{abstract}

\section{Introduction}

Over the last decade, the numbers of spectroscopic and photometric time-domain observations of stars have increased by many orders of magnitude. Single ground-based spectroscopic surveys now observe on the order of 10$^5$ - 10$^6$ stars. These surveys obtain measurements of stellar \teff, \logg\ and surface chemical compositions ([Fe/H], [X/Fe]). Such data provide insight into the large-scale formation of the Milky Way as well as test stellar physics on small scales \citep[e.g.][]{Freeman2002}.  Concurrently, high-cadence and long-baseline time-domain surveys have observed on the order of a few 10$^5$ stars and the TESS mission \citep{Ricker2014} will observe around 0.5 $\times$ 10$^6$ stars. Time-domain surveys, such as Corot \citep{Ridder2009} and Kepler \citep{Bedding2010}, which take high-cadence, precision photometry over a number of years, have produced critical information about stellar physics and interiors. This has been enabled by the precision measurements of \logg, radii and and mass -- which implies a subsequent stellar age \citep[e.g.][]{Chaplin:2014jf,2016MNRAS.455..987C,2018MNRAS.475.5487S}, which can be made using these data. The availability of large data volumes has motivated the development of new (and automated) tools and approaches to make parameter measurements from both spectra and multi-epoch photometry.

Using high-cadence stellar photometry, the asteroseismic parameters \numax\ \citep{Brown1991,Belkacem2011} and \deltanu\ \citep[][]{Ulrich1986} can be precisely measured from the frequency comb in the power spectrum. These parameters relate directly to the stellar mass through scaling relations \citep[e.g.][]{Kjeldsen1995,Stello2009,Huber2011}. Automated approaches applying Bayesian inference and neural network classification have had success in extracting these modes from the power spectrum \citep[e.g.][]{Davies2016, Lund2017, Janes2017}.

Given the stellar \teff, which is typically sourced from external photometry or spectroscopy, precise surface gravities can be determined from the measured \numax\ and  \deltanu\ using the scaling relations. Additionally, \logg\ can be inferred from the low frequency convective granulation signature (which stochastically drives the excitation of the higher frequency modes, see \citet{Houdek2015} for a recent review). The granulation signature is propagated to the observed signal as this is the activity arising from the top of the convection zone. The 8-hour Flicker method \citep{Bastien2013, Bastien2016, Cranmer2014}  uses this relation which arises between the 
\logg\ value and the time scale of the 
granulation \citep{Kallinger2016}. Surface gravity can similarly be inferred from the imprint in the autocorrelation function  (ACF) of the power spectrum \citep{Janes2017}. Recalibrating the 8-hour Flicker approach of \citet{Bastien2013} to correlate with the mean stellar density, this method has also shown to be relevant for planet characterization \citep{Tingley:2010ez,Kipping}. 

In the power spectrum, the granulation signal has a characteristic shape which is typically parameterised as a sum of power-laws centered on zero frequency. From the first proposed pure Lorentzian form by \citet{Harvey1985}, the adopted prescription for the granulation background has since taken different, slightly modified forms \citep[e.g.][]{Aigrain2003,Karoff2013,Kallinger2014}. Common to these parameterizations is the possibility of obtaining a characteristic time-scale and rms amplitude for the respective granulation components. The relation between these parameters and global seismic parameters is well established \citep{Kallinger2014}, and can be used to estimate global seismic parameters even when oscillations cannot be detected \citep{Campante2014}.  

Theoretical predictions have set the expectation that granulation should also contain information on \teff\ and be correlated with the asteroseismic parameters \citep[e.g.][]{Kjeldsen1995,Mathur,Samadi2013}. Given these theoretical expectations, we are motivated to test, empirically, what parameters we can derive from the time domain photometry and to what precision. Temperature is of particular interest to attempt to derive from the time-domain data. This is currently an external parameter input from spectroscopy, where possible (which provides higher precision estimates than photometry), into the $\numax$ and $\deltanu$ scaling relations \citep[e.g.][]{SilvaAguirre:2012du,Huber:2017fg}. Therefore, it would be useful to derive precise temperatures self-consistently from the multi-epoch photometric data. 

Complementary spectroscopy is delivering not only \teff\ but chemical compositions for a subset of the stars that have been observed with multi-epoch photometry. 
Combining the asteroseismic data with the spectroscopic data, where stars observed in both regimes, \citet{Martig2016} and \citet{Ness2016} were able to generate ages for $\approx$ 70,000 red giant stars in the \apogee\ survey across galactic radii 4--16 kpc, by learning from the Kepler observations where mass information resided in the \apogee\ stellar spectra. There is an opportunity to further build upon this data-driven approach, that enables us to learn how stellar labels correlate with stellar spectra, in the time-domain. In this work, we therefore take the data driven methodology as developed in \citet{Ness2015} of \tc\ and implement this approach on the Kepler time-domain photometry.  We choose four stellar parameters, or labels, of \teff, \logg, \numax\ and \deltanu\ to infer from the Kepler data for this first look at this prospect of data-driven inference. This is motivated by investigating: (i) if \teff\ can be recovered from the Kepler data, given that it is typically externally sourced  (ii) the precision to which labels can be recovered using a very simple data-driven model (iii)  the inference of \numax\ and \deltanu\ without using the frequency comb itself. We also look at the recoverability of stellar metallicity, \feh.

In Section 2 we present a summary of our data. In Section 3 we detail our method including all steps to reproduce our results, (with a link to the code). In  Section 4 we present our results and we discuss and conclude in Sections 5 and 6, respectively.

\section{Data}

We use the ${\sim}$4-year long-cadence Kepler data (85\% of our sample have a baseline of $>$ 3.5 years), downloading the unweighted power spectrum of 2000 red giant stars from the catalogue of \citet{Yu2018} from the KASOC database \citep{Handberg2014}.

These stars have a \teff\ and \logg\ available from \apogee, measured from H-band stellar spectra \citep{Majewski2017, P2014} and a \numax\ and \deltanu\ from the asteroseismic measurements of \citet{Yu2018}.

We divide these 2000 stars evenly up into a reference and test set of 1000 stars each. The reference set of stars is used to train our model (see Section 3) and the test set of stars is used to validate our model and test the precision of our results.

The uncertainty on the measured input parameters for the reference stars is $<$ 100K in \teff\ and $<$ 0.1 dex in \logg, from the spectroscopic measurements \citep{GP2016}. The asteroseismic parameters have typical uncertainties of  $<$ 1.6\% in \numax\ and $<$ 0.6\% in \deltanu\ \citep{Yu2018}. These stars have typical apparent magnitudes of H$\approx$10. 

\begin{figure}[]
\includegraphics[scale=0.45]{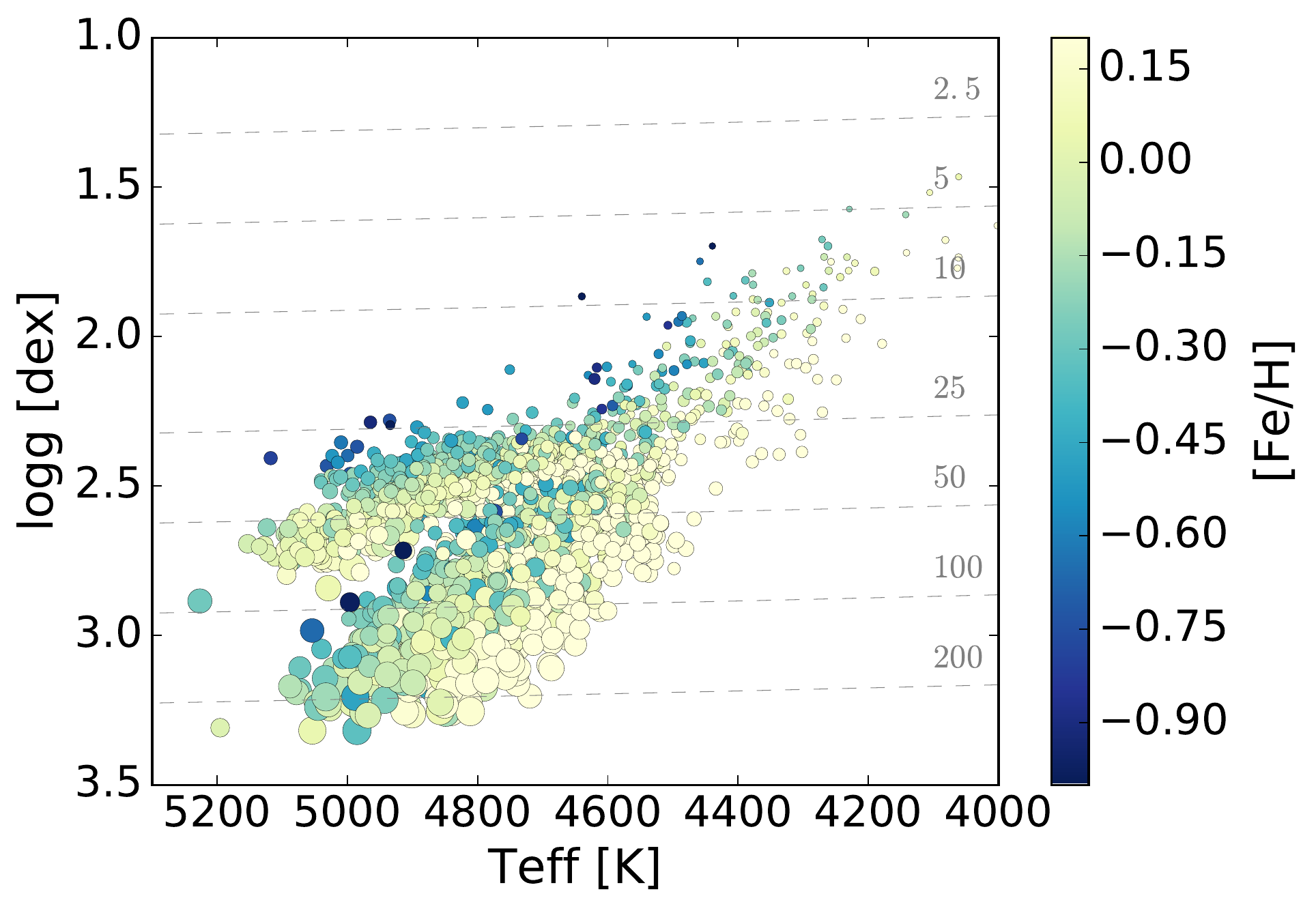} 
 \caption{The \teff-\logg\ plane of our training data, where the data is coloured by the \feh\ of the star and the size of the data points is the \deltanu, which ranges from \deltanu\ = 0.6 to 30 $\mu$Hz. The dashed lines show the iso-\numax\ contours, marked with their corresponding values in $\mu$Hz, following the relation from \citet{Campante2014}.  }
\label{fig:training}
\end{figure}

The \teff-\logg\ plane of the 1000 reference stars is shown in Figure \ref{fig:training}. In this figure, the marker size corresponds to the \deltanu\ value of the stars, with smallest \deltanu\ values corresponding to the lowest \logg\ values.  The range of parameters for these reference stars is: \teff = 3960 to 5226 K, \logg = 1.5 to 3.3 dex, \numax = 5 to 244 $\mu$Hz and \deltanu = 0.85 to 18 $\mu$Hz. 

\section{Method}

We use The Cannon almost directly as first outlined in \citet{Ness2015}. \tc\ is a data-driven approach and relies on a reference set of stars with known `labels' that describe the stellar flux. This reference set of stars and their associated labels is used to build a model. The model, which we assert as a simple polynomial function of the labels at each value (or pixel) of our amplitude function (described below), is then used to return the stellar labels for new stars at test time \citep[see][for some examples of applications of this approach]{Ness2018}. 

In this case, rather than infer our labels from the spectroscopic data \citep[as per][]{Ness2015}, we infer labels from the autocorrelation function (described below) of the Kepler power spectra. Here we are first interested in four labels: \teff, \logg, \numax\ and \deltanu. 

\subsection{Processing of the Kepler data for \tc}

(i) We download the unweighted power spectrum from the KASOC database for our 2000 stars.

\begin{figure}[]
\centering
\includegraphics[scale=0.45]
{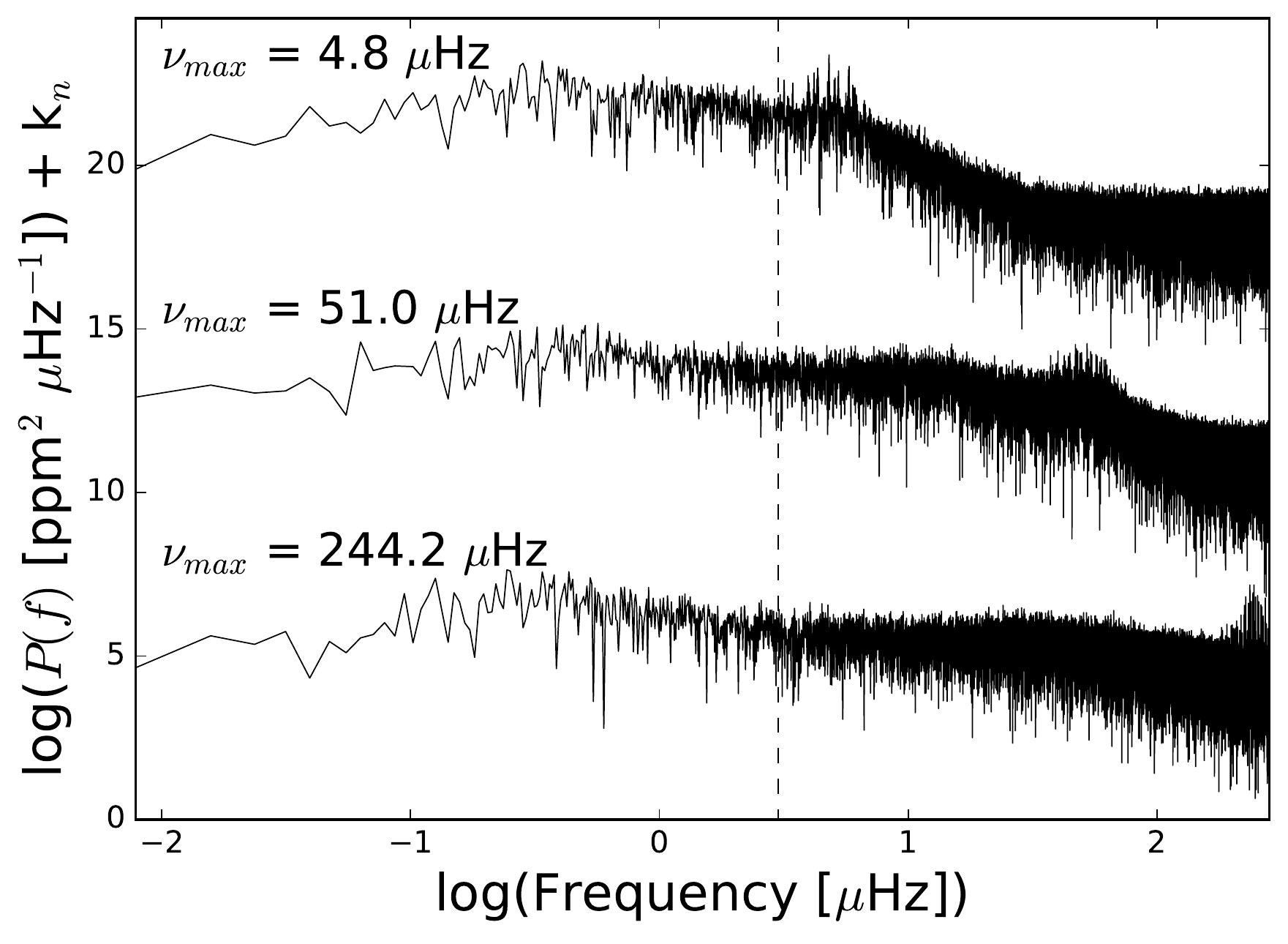} 
 \caption{Sample smoothed unweighted power spectrum of three stars from our set of 2000 stars, spanning the range in for $\numax$ of these objects. Each spectrum is offset by a constant value, k, to separate the flux on the y-axis. The dashed line indicates the value of 3$\mu$Hz, which is the lower bound of the frequency range used to build our model. The upper frequency bound of our model is the upper limit in the figure.}
\label{fig:3stars}
\end{figure}

(ii) The first processing step is to interpolate the data to a common grid. We select frequencies 3-270$\mu$Hz and interpolate in 0.009$\mu$Hz steps. We found empirically that slightly improved precision was obtained by removing the first few $\mu$Hz. The first few $\mu$Hz will typically be affected by the light curve correction, and may contain signal from stellar activity -- it is therefore not expected that the lowest frequencies correlate cleanly with stellar parameters. Our upper limit of 270$\mu$Hz is slightly higher than the largest \numax\ value in our stellar sample, and near the Nyquist frequency limit, at ${\sim}$283$\mu$Hz of the KASOC generated power spectra. Three examples of the downloaded power spectra, for stars across the range of $\numax$ values are shown in Figure \ref{fig:3stars}.

For \tc\ as presented in \citet{Ness2015} to be applied to the time-domain data, the following must hold true: (a) stars with the same labels must have the same spectra and  (b) stellar flux at a given wavelength must vary smoothly with stellar labels. However, the power spectra from which the asteroseismic parameters \numax\ and \deltanu\ are typically determined \citep[e.g.][]{Yu2018} do not satisfy condition (b) of \tc. In the power spectrum, the changes with the labels occur sharply and not smoothly as a function of amplitude, because oscillation modes appear as extremely narrow peaks in the power spectrum. Since oscillation modes of giant stars are so coherent and therefore produce narrow peaks in Fourier space, slight shifts in the positions these peaks produce very large changes in power at a given frequency. We must therefore work with the data in the time domain, where the labels correlation if present is a function of amplitude (in part, they may also be a function of time). 

(iii) Our second processing step is then to transform the power spectrum in the frequency domain, $f$, from the KASOC database back into the time domain, $t$, using the inverse Fourier transform of the power spectrum function $g(f)$ (using the \texttt{numpy} function \texttt{ifft} in \texttt{python}). The discrete inverse fourier transform of our power spectrum $P(f)$ is then;

\begin{equation}
  \label{ift_continuous}
   g(t) = \frac{1}{\sqrt N }\sum_{f=0}^{N-1}  
e^{( 2 \pi i t f / N) } P(f) 
\end{equation}

Where N = FFT size, $t$ = time step, $f$ = frequency step, for  f =  0,1..N-1.

This inverse Fourier transformation step essentially re-factors  information in the data to be smoothly varying with amplitude. The signal, which is sparse in the frequency domain, becomes dense in the time-domain.  In the time-domain, the information at each frequency point is distributed across all pixels. This transformed signal is the autocorrelation function (ACF) of the time-series data. We could work with the light curve data directly, however the ACF has the advantage that it quantifies the time correlation of the observed brightness fluctuations \citep[see][]{Kallinger2016} and is a continuous function of the generated power spectrum.   The autocorrelation essentially characterises the amplitude of periodic signals in the data, so will quantify any regular surface brightness fluctuations, which we can then test for label correlation.  The ACF has been very effective in automated determination of the rotation period of stars and has been shown to be robust to systematic effects in the data \citep{McQ2013}. 

We take the absolute value of the ACF ($|g(t)|$), which we find correlates with our labels. We work with the logarithm of this amplitude due to the large range in its magnitude:
\begin{eqnarray}
G(t) = \ln (|g(t)|)\,
\end{eqnarray}
this enables us to use the a simple quadratic model, as per our spectroscopic analyses in \citet{Ness2015}. 

\begin{figure*}[]
\centering
\includegraphics[scale=0.45]
{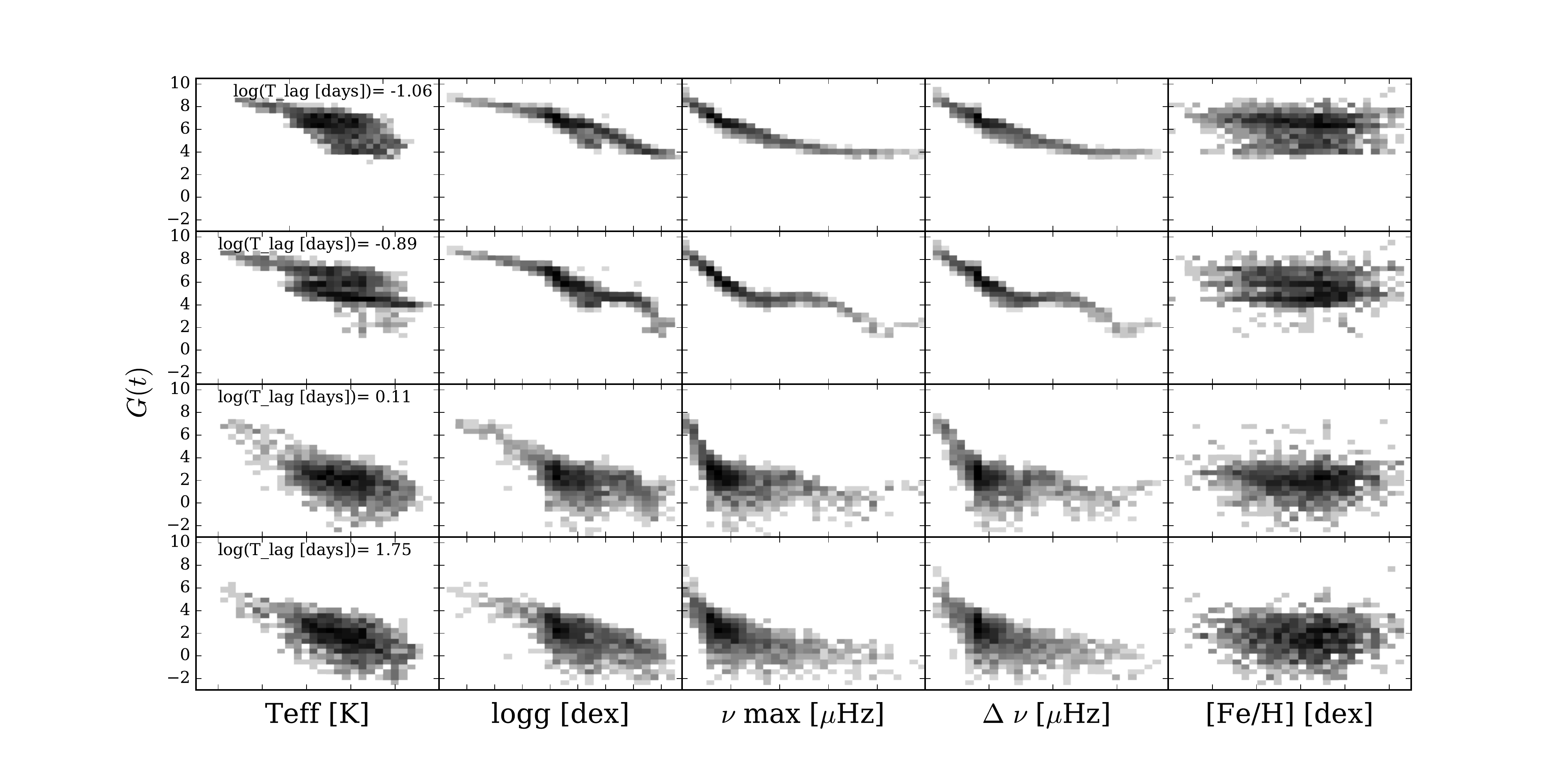} 
 \caption{This Figure shows the information content in single pixels. Each row corresponds to a single pixel with the period indicated in the first sub-panel of each row. These pixels correspond to the arrows marking their location in the Figure \ref{fig:coeffs}. The $G(t)$ amplitude  is shown on the y-axis and and the four labels we infer, of \teff, \logg, \numax, and \deltanu\ is shown for each pixel on the x-axis. The \feh\ result is also shown in the last panel, indicating there is no correlation between metallicity and the amplitude $G(t)$.  This Figure illustrates why \tc\ can work to infer the four labels of \teff, \logg, \numax\ and \deltanu\ using a polynomial model that relates the function $G(t)$ to the labels. }
\label{fig:onepix}
\end{figure*}

An example of the label correlation for the \teff, \logg, \numax, \deltanu\ and \feh\ labels of the reference stars, in four pixels of the ACF amplitude, $G(t)$, is shown in Figure \ref{fig:onepix}. There are clear correlations between the \teff, \logg, \numax, \deltanu\ labels and individual pixels. In addition to showing the correlation between the amplitude and the labels we infer, we also show the metallicity, \feh\ of the stars in the last panel of Figure \ref{fig:onepix}. This demonstrates that there is no metallicity information in these pixels (nor any other pixels in the ACF).  Subsequently, when added as a label to the model, metallicity can not be inferred from the data.

\subsection{Generating \tc's model from the ACF amplitude}

With the conditions for \tc\ now satisfied and a clear correlation between the ACF amplitude and labels established (and shown for a few sample single pixels in Figure \ref{fig:onepix}), we can now build our model from the reference objects for training. 

We take our $n=1000$ reference objects with their known labels $\starlabelvec_n$. The spectral model is characterized by a coefficient vector $\set{\theta}_t$ that allows the prediction of the function $G(t)$ amplitude which is $G_{n_t}$  at every time step, t, of the ACF  for a given label vector: 

\begin{eqnarray}
G_{n_t} &=&
function(\starlabelvec_n |  \set{\theta}_t) + \mbox{noise}
\label{eq:specmodel}\quad 
\end{eqnarray}

This relates stellar labels $\starlabelvec_n$ to the ACF function amplitude $G_{n_t}$ at each time step. 

This model leads to the single-pixel log-likelihood function:

\begin{align}
\ln p(G_{n_t}\given\set{\theta}^T_t, \starlabelvec_n, s_t^2) =
 &-\frac{1}{2}\,\frac{[G_{nt} - \set{\theta}^T_t \cdot \starlabelvec_n]^2}{s_t^2 + \frac{\sigma_{nt}^2}{g_{nt}}} \\
 &-\frac{1}{2}\,\ln(s_t^2 + \frac{\sigma_{nt}^2}{g_{nt}}) \notag
\label{eq:like}\quad.
\end{align}

The noise term here is an rms combination of the
associated uncertainty variance $\sigma_{n}^2$ of each of the pixels of the
ACF from finite photon counts and instrumental effects and the
intrinsic variance or scatter of the model at each time-lag pixel of
the fit, $s^2$. Here, as our data is transformed to a logarithmic amplitude, a term of $1/g_{nt}$ appears in the denominator as the propagated amplitude error. Deviating from the original implementation of \tc, we do not include any error model for $\sigma_n$ for our data. We did tests incorporating the errors, sourced and propagated to the ACF from the original light curve data, however they did not improve (but, rather, marginally degraded) the precision of our inferred labels.

At training time, the coefficient vector is solved for (at every pixel, or time-lag step of the ACF). In the functional form of equation (3), there are fifteen coefficients at each pixel, such that $l_n$ = [1, \teff, \logg, \numax, \deltanu, (\teff\ $\cdot$ \logg), (\teff\ $\cdot$ \ \numax), (\teff\ $\cdot$ \deltanu), (\logg $\cdot$ \numax), (\logg\ $\cdot$ \deltanu), (\numax $\cdot$ \deltanu), \teff$^2$, \logg$^2$, \numax$^2$, \deltanu$^2$].

These correspond to the terms of a quadratic model {$\theta_t$} (i.e. the single label terms, squared label terms and cross label products of a second order model) as well as the scatter, {$s_t^2$} in the noise term. 

\begin{figure*}[]
\centering
\includegraphics[scale=0.45]
{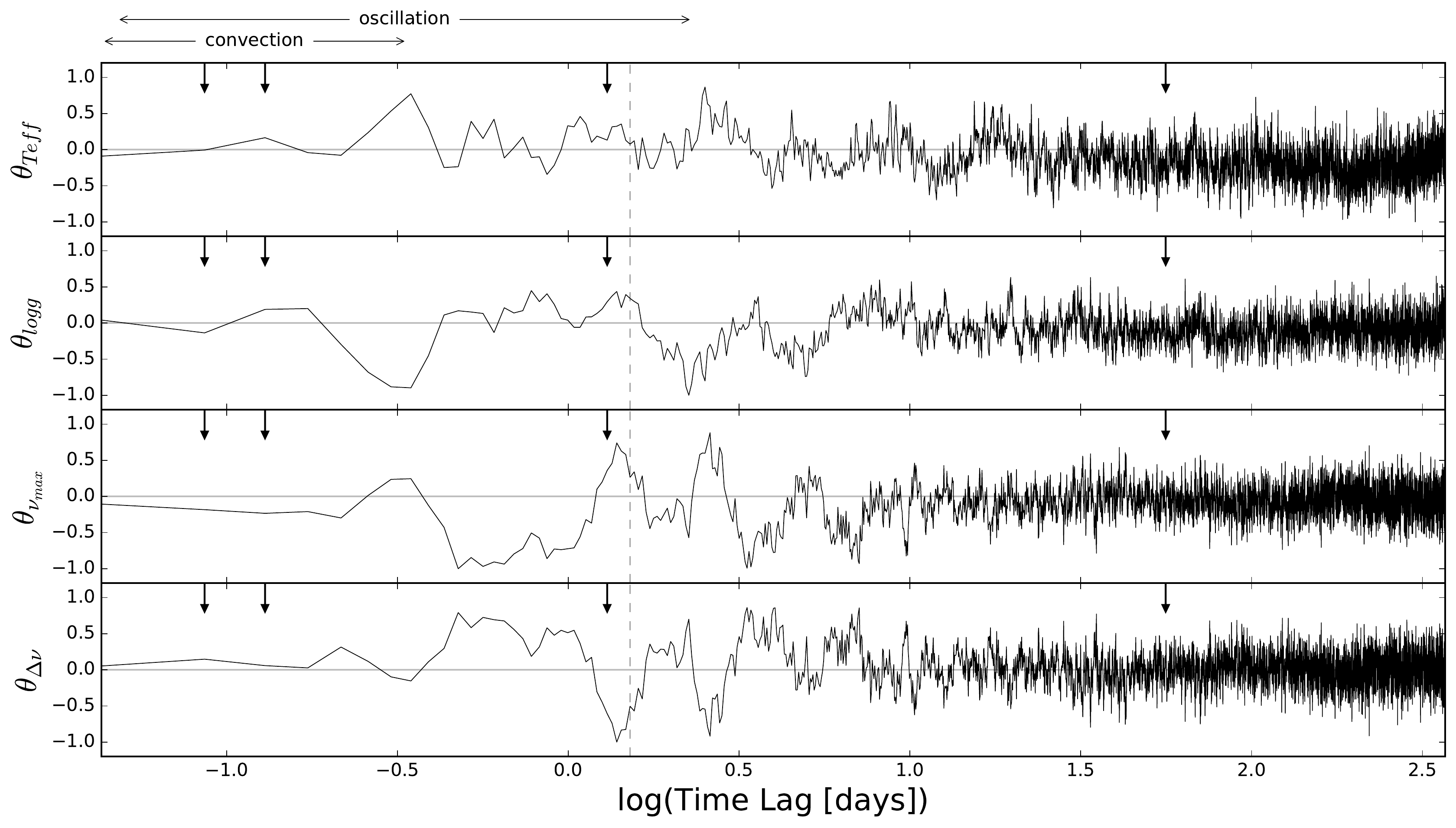}
 \caption{The first order coefficients of our (polynomial) model corresponding to our four inferred labels, scaled to the absolute maximum value of each coefficient. The time-lag is shown in logarithm scaling in order to better show the structure present at the smallest time-lag values. This reveals that \numax\ and \deltanu\ are oppositely correlated and the \logg\ information is coming from regions that are not entirely correlated with the two asteroseismic parameters. The dashed line represents the cutoff period for the model to infer \logg, \numax\ and \deltanu. The full extent of the period shown on the x-axis is the full model used to infer \teff. The convection zone and acoustic oscillation zone (which is stochastically driven by turbulent convection) are indicated, (where the convection zone is the region used by the 8-hour Flicker method). The four arrows within the sub-panels correspond to the pixels selected to demonstrate the correlation of the labels and the amplitude, shown in Figure \ref{fig:onepix}.}
\label{fig:coeffs}
\end{figure*}

\tc's first order coefficients from the training step are shown in Figure \ref{fig:coeffs}.  This time-lag range of {0.045 $<$ \tlag\ $<$ 370 days} (-1.3 $<$ $\log (\tlag$ (days)) $<$ 2.6) was determined from empirical tests to be the range we needed for our model, so as to infer \teff\ to the highest precision. Our model extends to a shorter time-lag range of 0.045 $<$ \tlag\ $<$ 34 days (-1.3 $<$ $\log (\tlag$ (days)) $<$ 1.5) to infer the other three labels, again the necessary range for the best precision we could achieve. These first order coefficients that are shown are scaled to their respective maximum absolute values, so the anti-correlations and relative behaviors can be compared. Large positive or negative values indicate where the ACF amplitude is most sensitive to the label. Note the largest amplitudes are in the lower time-lag regime. For the \teff\ label there is an overall continuum shape across the full time-lag period range not seen for the other labels (an overall decrease at $\log( \tlag$ (days)) $\approx$ 2. 

The coefficients quantify where the information with respect to the labels is contained in the data, as a function of the time-lag. This  can be used to compare to theoretical expectations and also to search for the information content with respect to any other labels (see the Discussion, we attempted to learn metallicity and found no [Fe/H] correlation). There are interesting features in the first order coefficients. For example, we see two clear peaks in the $\numax$ and $\deltanu$ coefficients that are anti-correlated. These peaks may be a consequence of the different granulation components (from different convective scales). These different granulation components have different characteristic amplitudes and timescales (although are fixed with respect to each other for different stars, i.e., you always find the same ratio between timescales and amplitudes). Therefore, it would make sense that the amplitude sensitivity would vary as these different components are encountered in the time-log representation. Figure \ref{fig:coeffs} also shows that the \logg\ information is not directly correlated with that of \numax\ and \deltanu. An investigation of the behaviour of the coefficients is worthy of follow up, but a detailed explanation is beyond the scope of this paper. 

We select four time-lag pixels from Figure \ref{fig:coeffs}, which are indicated by the arrows in this Figure. Figure \ref{fig:onepix} shows the amplitude of the ACF of the reference stars at these for these four pixels, as a function of their \teff, \logg, \numax, \deltanu\ and \feh\ values. These four pixels are indicated in Figure \ref{fig:coeffs} by the arrows.  All pixels show correlations in \teff, \logg, \numax\ and \deltanu, but there is no correlation with \feh. Figure \ref{fig:onepix} demonstrates that the different time-lag values show different forms of correlation with the labels (which is quantified overall, in the coefficients). For example, the first time-lag pixel we select, at $\log(\tlag$ (day)) = -1.06 (or $\sim$ 2 hours), has the smallest first order coefficient of \teff, of the four pixels. Figure \ref{fig:onepix} shows that this pixel also has the least correlated temperature dependence of the four pixels. This particular value of the time-lag also shows relatively small first-order coefficients of \logg, \numax\ and \deltanu\  (compared to other time-lag values). However, we still see striking correlations in these labels at this pixel. We emphasize, that we show here only part of \tc's model; e.g. the squared terms also quantify and contain the correlation terms. Therefore, it is important in any quantified analysis of the information content at a given pixel to consider the model as a whole. Here we select the first order coefficients as a qualitative illustration.

At the inference step, or the test step, we use the coefficients solved for in training as inputs in the same equation as in the training. Thus, we learn the four labels given the test data \citep[see][for details]{Ness2015}. 

Our code can be found at the github repository: https://github.com/mkness/ACFCannon

\section{Results}

\begin{figure*}
\centering
\includegraphics[scale=0.45]
{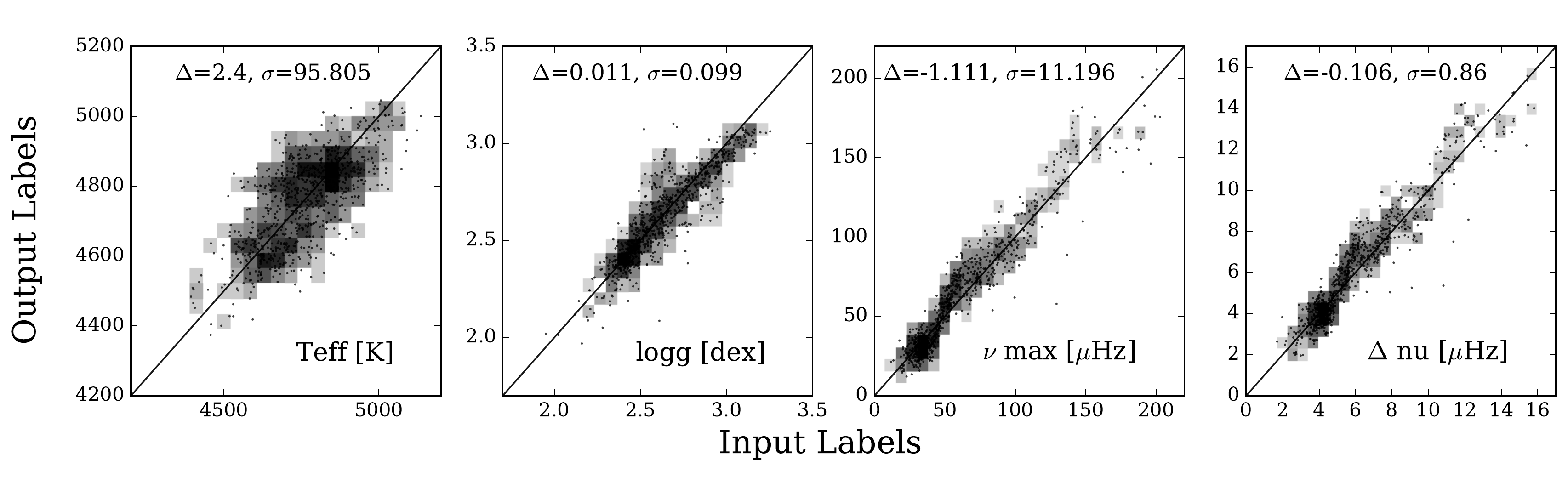} 
 \caption{The performance of our combined model on the test stars (that were not used for training of the model). We remove $\approx$ 150 test stars with poor model fits using the $\chi^2$ selection of 150 $>$ $\chi^2$ $>$ 1000 (calculated over the 780 pixels used for our model for \logg, \numax\ and \deltanu.). Output labels represent our results and input labels are from \apogee\ (for \teff\ and \logg) and \citet{Yu2018} (For \numax\ and \deltanu). We can infer \teff\ precise to $<$ 100 K , \logg\ precise to $<$ 0.10 dex, \numax\ precise to $<$ 10.5 $\mu$Hz and \deltanu\ precise to $<$ 0.9 $\mu$Hz. Periods of P$_{\rm acf}$ $<$ 370 days are used for the \teff\ determination and periods of P$_{\rm acf}$ $<$ 34 for the other labels.}
\label{fig:validation}
\end{figure*}

\begin{figure}
\centering
\includegraphics[scale=0.45]
{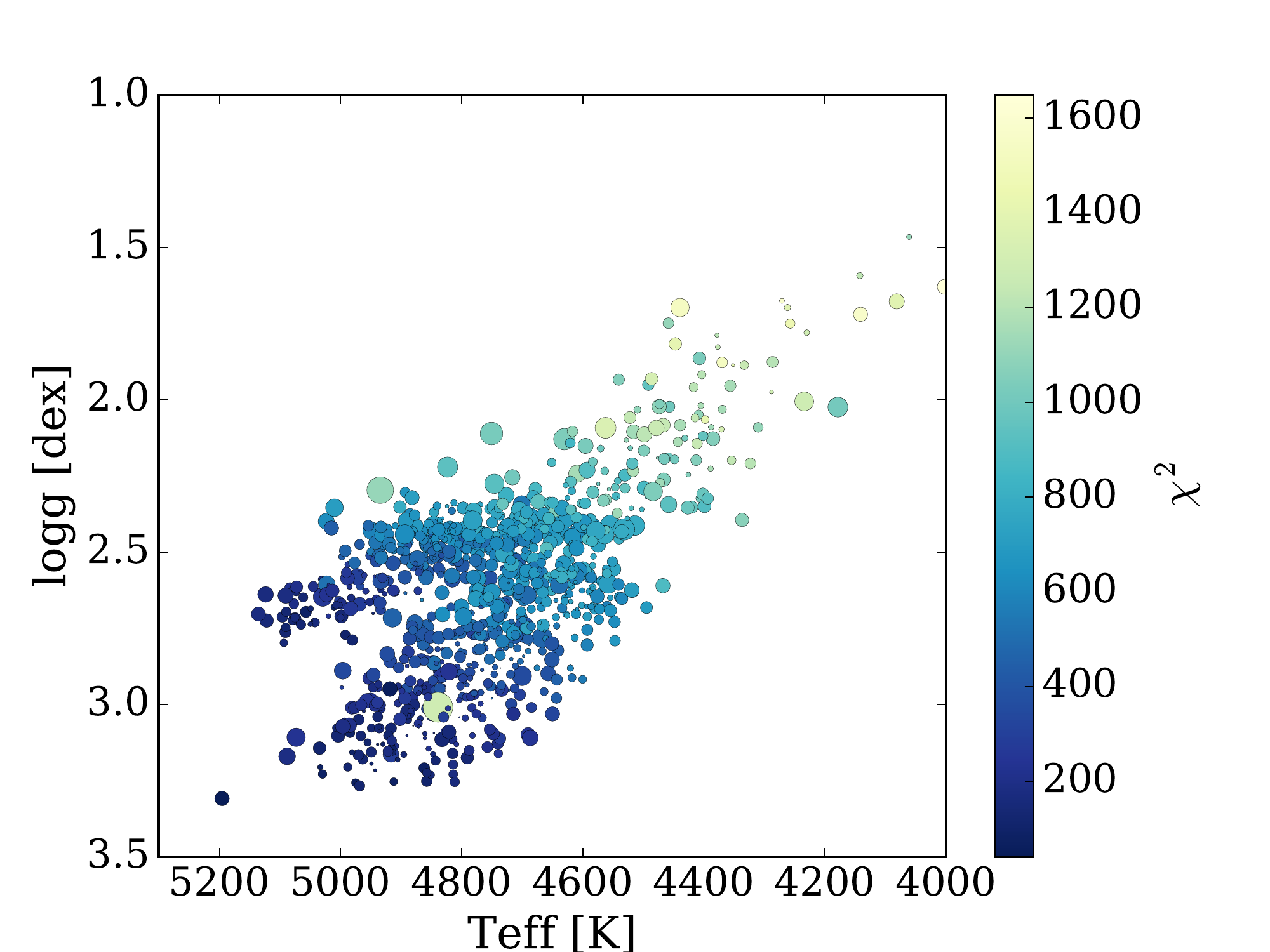} 
 \caption{The 1000 test stars, coloured by the $\chi^2$ of the model fit. The marker size corresponds to the absolute difference of the \logg\ value between the inferred (output) value and input (reference) value, for that star.}
\label{fig:chi2}
\end{figure}

We show the performance of \tc\ at test time using the 1000 red giants which have \textit{not} been used in our reference set to create our model. Their known external labels compared to \tc's model are shown in Figure \ref{fig:validation}. For this Figure we make a $\chi^2$ cut to remove poor fits to the model, selecting stars with 150 $<$ $\chi^2$ $<$ 1000, after which 850 stars remain of the original 1000. Figure \ref{fig:chi2} demonstrates that it is the regions that are sparsely populated with training data where the $\chi^2$ values are large. However, we also find that what is driving the higher $\chi^2$ at lower \logg\ values is the removal of the first few $\mu$Hz of the power spectrum for our data modeling. The $\numax$ value is highly correlated with the \logg\ shown in this Figure. For the highest $\numax$ values in our sample, the removal of the first 3$\mu$Hz of the power spectrum removes more of the granulation background containing information correlated with our labels. The 3 $\mu$Hz cut is effectively tuned for this subset of parameter space, and therefore optimal around the median of the labels. Although our model works to the precision of spectroscopy for \logg\ and \teff\ label inference, this is indicative that we could come up with a better model, that works across a broader \teff-\logg\ parameter space.  

This Figure shows that the uncertainty on the \teff\ and \logg\ is on the order of that of the input labels themselves (which is $<$ 100K and $<$ 0.1 dex respectively). The \numax\ and \deltanu\ uncertainties (of $\lesssim$ 15\%) are however significantly higher than the input uncertainties on these labels (of about 1.6\% and 0.6\% respectively), presumably as these are not being measured from the discrete modes themselves but from the long-period granulation signature: we note again that these asteroseismic parameters are being inferred from the mode-exciting convective granulation at an ACF time-lag of \tlag\ $<$ 34 days. While not providing precise measurements from the convection pattern themselves, these can still provide a useful starting point for a more detailed measurement from the mode frequencies \citep[e.g.][]{Huber2010,Hekker2011,Mosser2012,Stello2017,Serenelli2017}. That the information is available from the granulation signal agrees with theoretical expectations: the acoustic modes from which the asteroseismic parameters are defined are stochastically driven by the turbulent convection in the outer regions of the star. The observed granulation signal is the surface manifestation of this turbulent convection, hence a causal relation can be expected between this signal and the asteroseismic parameters.

\begin{figure*}
\centering
\includegraphics[scale=0.45]{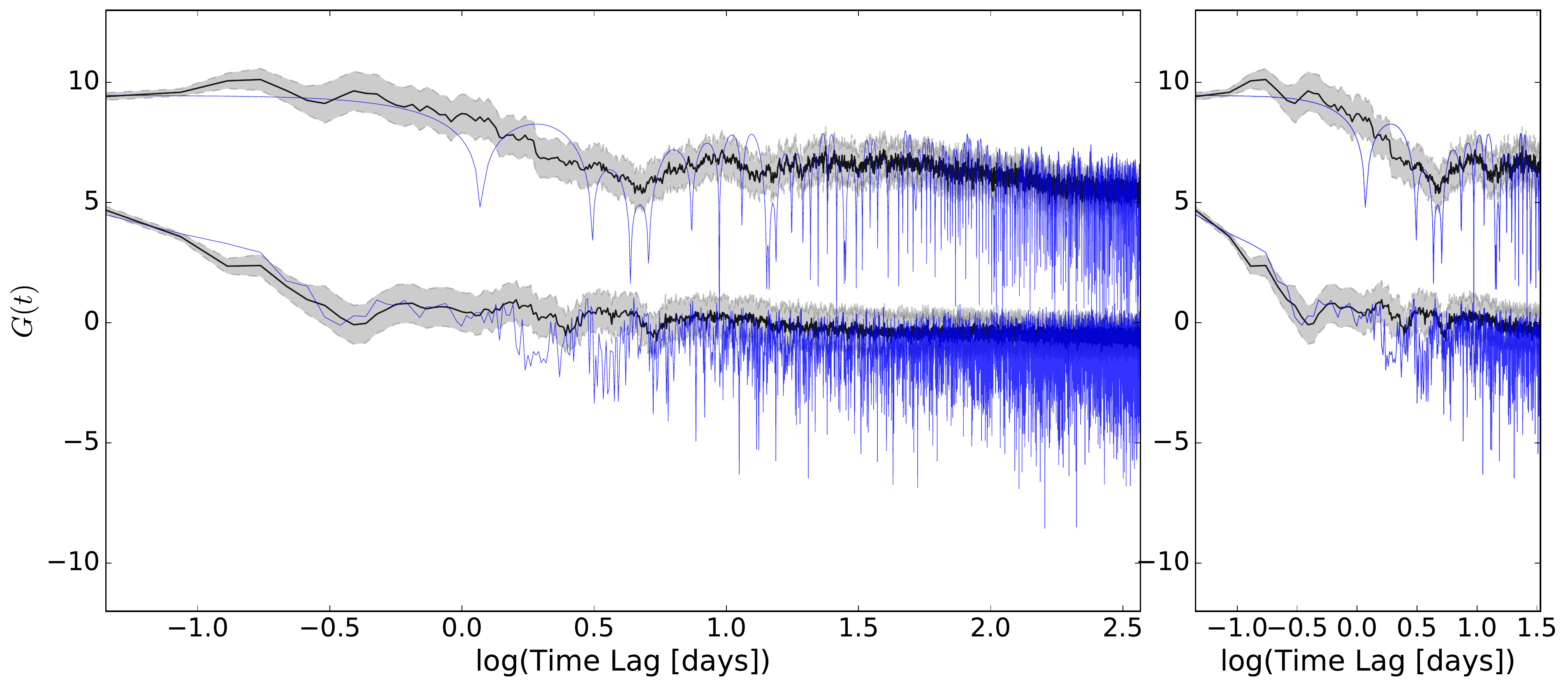} 
 \caption{Two sample stars with dissimilar parameters showing the data (blue) and best fit model from \tc\ (black). The scatter term representing the uncertainty of the model shown in the gray shaded regions around the black model line. The panel at left shown the full region of the model used to infer \teff\ and the right hand panel shows the region used to infer \logg, \numax\ and \deltanu. The model reproduces the overall amplitude response of the data and does not capture the detailed structure across period. Nevertheless, this overall amplitude information is sufficient to determine the four labels at high precision, in fact on the order of the input label precision for \teff\ and \logg. The respective parameters of the two stars are
\teff\ =4000 and 5550K, \logg\ = 1.6 and 3.3 dex, \deltanu\ = 1.0 and 3.3 $\mu$Hz and  \numax\ = 5.5 and 104.7 $\mu$Hz, respectively. $\chi^2$ = 9991 and 6811 for the two stars, respectively, across 8500 pixels.}
\label{fig:twostars}
\end{figure*}

In Figure \ref{fig:twostars} we show \tc's best fit model to two example stars that have different labels. This figure shows that \tc's model is learning not the detailed response of the ACF but the overall amplitude pattern.  The model represents the underlying noise-free limit-spectrum of the data. Under the assumption that the stars are drawn from some underlying true distribution, the model is analogous to a de-noised version of the data. The high value of our scatter term implies that our uncertainties can be reduced with a more sophisticated choice of model (beyond a polynomial function) -- if indeed additional information is physically present. An advantage and choice motivating the polynomial form here is that it is interpretable (see the Discussion) and very fast to test for the information content in the data so provides a very good starting point for this data-driven analysis.

Empirical testing demonstrated that no precision gain was obtained by going to a time-lag beyond 370 days in the ACF amplitude $G(t)$ for inferring \teff. The precision was in fact marginally higher when restricting to a time-lag of $\leq$ 34 days for learning the other three labels. We note again that we also found slightly higher precision from removing the first few $\mu$Hz of the data before calculating the ACF, meaning our minimum period is 0.045 days.

\section{Discussion}

We have taken a first look at data-driven inference of parameters using a simple polynomial description of the time-lag amplitude function, the ACF of the Kepler power spectra. We determine the ACF directly from the inverse Fourier transform of the KASOC power spectrum. Beyond our cuts in the initial frequency selection of the power spectrum (3-270 $\mu$Hz) and then the time-lag of the generated ACF, we require no additional processing nor filtering of the data in our approach.  

We find that it is possible to infer \teff\ and \logg, at the precision of spectroscopy and relatively imprecise \numax\ and \deltanu\ for red giant stars. We attempted to infer the \feh; this label is available from the \apogee\ spectroscopy for our stars. However, this label failed and on inspection, no pixels correlated with \feh. Therefore, contrary to the findings by \citet{Corsaro2017}, we find that there is no information with respect to \feh\ in the granulation signal from the \textsl{Kepler} multi-epoch photometry. We note that corrections to scaling relations between \deltanu\ and fundamental stellar parameters include both \teff\ and \feh\ \citep{White2011,Guggenberger2016,Sharma2016}. Furthermore, \citet{Viani2017} showed, in the case of \numax, a dependence on mean molecular weight. While we do not find the signature in the ACF amplitude, this does indicate that a \feh\ dependence might be expected, as also suggested by 3D hydrodynamical simulations of convection \citep[][]{Collet2007,Ludwig2016}.

The data-driven approach has the advantage that potential unknown  influences on the adopted labels are automatically taken into account. An example could be a potential effect from magnetic suppression of convection as discussed in \citet{Cranmer2014} based on the theoretical scaling relation of \citet{Samadi2013}. \citet{Cranmer2014} adopted a parametric correction to the granulation variability as a function of \teff\; in the data-driven approach such a correlation will be automatically included without the need for specifying a parametric relationship.

We emphasize that this work is a first look at this data-driven approach. This generalized approach can be extended directly to other evolutionary states, including main sequence and turn-off stars. Additional labels may be able to be inferred, such as stellar radius and mass directly, given a more sophisticated model. The polynomial function is only suitable for labels that show the correlations similar to those seen in Figure \ref{fig:onepix}. 

A primary original motivation of \tc\ was to enable all spectroscopic surveys to be put directly on the same label (stellar parameter and abundance) scale, given stars in common between different surveys. Here an internal determination of temperature is particularly relevant to this goal, in that all time domain data can be placed on a common temperature scale, which is important for consistent estimates of stellar mass, radii and age \citep[e.g.][]{Epstein2014,Tayar:2017du,2018MNRAS.475.5487S}.

The time-lag periods of the ACF that we selected for the inference of parameters were determined empirically, as marginally providing the highest precision parameters. We therefore want to first see if we can now interpret these ranges in the context of expectations from theoretical models. \citet{Mathur} show that the period is theoretically a function of the \numax; $P_{\rm gran}$ $\approx$ $\nu_{\rm max}^{−1.90}$, which is consistent with the theoretical predictions. This value ties in very well with the range of time-lag that we find is required for the red giants. Furthermore, \citet{Mathur} find that granulation timescales of stars
that belong to the red clump are similar,  while the timescales of stars in the red giant branch are spread across a wider range. We expect this is why in Figure \ref{fig:validation} we do not resolve the \logg\ of the red clump as well as the red giant (see the pile up in the vertical direction at the clump logg around 2.5). Our lower precision on \numax\ corresponds to the expected theoretical precision on the granulation power from which we derive this value, estimated in \citet{Mathur} to be $\sigma_{\rm gran}$ = 8.6\%. 
Figures \ref{fig:onepix} and \ref{fig:coeffs} demonstrate the information from the brightness fluctuations is present across a multitude of individual amplitude values of the data, with different correlations at different pixels, that ultimately tie back to the convective patterns of the stars. This highlights that using a model of each data point, more information can be captured than in scaling relations, which are essentially integrated properties of the signal.

\section{Conclusion}

Our paper is a first investigation into a data-driven approach for the inference of parameters from multi-epoch spectra. Using the long-cadence observations of 2000 Kepler red giant stars,  we found we could infer \teff\ and \logg\ to the same precision as spectroscopy ($<$ 100 K and $<$ 0.1 dex, respectively). Furthermore, we recover the asteroseismic parameters \numax\ and \deltanu, precise to $<$ 11 $\mu$Hz and $<$ 0.9 $\mu$Hz, respectively (approx $<$ 15\%). We do not infer these four parameters from the frequency comb of the power spectra, but from the light curve ACF time-lag amplitudes. These amplitudes quantify the convection driven brightness variations that excite the high-order resonant acoustic modes of the star \citep[e.g.][]{Goldreich1994}. We attempted to learn \feh, but found no indication of information on this parameter in the multi-epoch data. 

The inference of \teff\ directly from the time-domain data enables a consistent temperature scale for stars within and also between surveys (assuming stars observed in common between the time-domain surveys). This homogeneous temperature scale enables the stars to be placed on a consistent stellar radii and mass scale. Furthermore, although simple, our polynomial model enables an empirical quantification of the data content with respect to derived parameters. Thus, enabling comparisons to theoretical expectations of the convective granulation. The method we demonstrate here on the red giant stars using the \textsl{Kepler} data is readily transferable to other evolutionary states, additional labels and future surveys. A fast delivery of \teff\ and \logg\ values and estimates of global asteroseismic parameters for a large number of stars will be extremely valuable for the recently launched TESS mission \citep[][]{Ricker2014}, which will provide light curves for thousands of red giants across the entire sky \citep{Lund2017b}.

\section*{Acknowledgements}

This project was advanced, in part, during the 2018 tess.ninja workshop, hosted by the Flatiron Institute, New York. We thank Guy Davies for useful discussion. 

Funding for the Stellar Astrophysics Centre is provided by The Danish National Research Foundation (Grant agreement no. DNRF106). VSA acknowledges support from VILLUM FONDEN (research grant 10118). MNL acknowledges support from the ESA-PRODEX programme.

\end{document}